\documentstyle[aps,preprint,eqsecnum,floats,epsf]{revtex}
\tighten \draft
\begin{document}
\preprint{}

\title{An Exactly Solvable Model of Interacting Bosons}

\author{A.~B. Balantekin\thanks{Electronic address: {\tt
      baha@nucth.physics.wisc.edu}}}
\address{Department of Physics, University of Wisconsin;\\  Madison,
Wisconsin 53706 USA}

\author{T.~Dereli\thanks{Electronic address: {\tt tdereli@ku.edu.tr}}}
\address{Department of Physics, Ko\c{c} University;\\
34450 Sar{\i}yer-\.{I}stanbul, Turkey}

\author{Y. Pehlivan\thanks{Electronic address: {\tt yamac@gursey.gov.tr}}} 
\address{Department of Physics , Middle East Technical University;\\
06531 Ankara, Turkey}

\maketitle

\begin{abstract}
We introduce a class of exactly solvable boson models.
We give explicit analytic expressions for energy eigenvalues and
eigenvectors for an sd-boson Hamiltonian, which is related to the
$SO(6)$ chain of the Interacting Boson Model Hamiltonian.
\end{abstract}

\section{Introduction}

A number of complex quantum systems can be described algebraically
using bosonic degrees of freedom. For example even-even nuclei can be
described in terms of a system of one boson with angular momentum
$L=0$, called $s$, and five components of another boson with angular
momentum $L=2$, called $d_{\mu}, \mu = 0, \pm 1, \pm 2$. This model,
called Interacting Boson Model, has been very successful in
describing the low-lying collective states of even-even nuclei
\cite{Armia:1976ky,o6} A similar boson model, using an angular
momentum $L=1$ ($p$) boson in addition to the $s$-boson has been
used to describe the structure of molecules \cite{molecules} and
hadrons \cite{hadrons}. The s-d boson model has an $SU(6)$
symmetry and the s-p boson model has an $SU(4)$ symmetry.
Analytic solutions of these models are given whenever the
Hamiltonians can be written in terms of the Casimir operators
of the $SU(6)$ (or $SU(4)$) group and its subgroups. In particular 
for the $SO(6)$ chain the Hamiltonian  contains the boson pairing
operator.

Since the boson creation operators $d^{\dagger}_{\mu}$ are spherical
tensors and the annihilation operators can be written as spherical
tensors after the introduction of a suitable phase
(i.e.  $ {\tilde d}_{\mu} = (-1)^{\mu} d_{-\mu}$)
we will use the dot product notation of spherical tensors, $(\cdot)$,
below. It is easy to show that the three operators
\begin{equation}
\label{1}
S^+ = \frac{1}{2} (d^\dagger \cdot d^{\dagger}),
\end{equation}
\begin{equation}
\label{2}
S^- = \frac{1}{2} ({\tilde d} \cdot {\tilde d}),
\end{equation}
and
\begin{equation}
\label{3}
S^0 = \frac{1}{4} \sum_{\mu} (d^\dagger_{\mu}  d_{\mu} + d_{\mu}
d^{\dagger}_{\mu}),
\end{equation}
generate the d-boson
quasi-spin group $SU(1,1)$. This group can also be represented
using only the s- or p-bosons. The eigenvalues of the operator 
$S^0$ are given by $(n/2)+(1+\alpha)/4$, where $n$ is the total
number of the bosons and
the quantity $\alpha$ is 0, 1, and 2 for the s-,p-, and d-bosons,
respectively. For the $SO(6)$ symmetry chain one needs to include
both the s- and d- bosons in the realization of the algebra:
\begin{equation}
\label{4}
S^+ = \frac{1}{2} [(d^{\dagger} \cdot d^{\dagger}) - (s^{\dagger}
s^{\dagger})],
\end{equation}
\begin{equation}
\label{5}
S^- = \frac{1}{2} [({\tilde d} \cdot {\tilde d} ) - (ss)],
\end{equation}
and
\begin{equation}
\label{6} S^0 = \frac{1}{4} \left[\sum_{\mu} (d^\dagger_{\mu}
d_{\mu} + d_{\mu} d^{\dagger}_{\mu}) + (s^{\dagger} s + s
s^{\dagger}) \right] .
\end{equation}
The pairing operator of the $SO(6)$ limit is then $P_6 = S^+S^-$, where
$S^{\pm}$ are given by Eqs. (\ref{4}) and (\ref{5}).

Our goal in this paper is to consider a system of s- and d-bosons
near, but not necessarily at the $SO(6)$ limit. It is worth to point
out that similar efforts were undertaken by others. In Ref. \cite{pan1}
the $U(5)$ to $SU(3)$ transition in the interacting boson model was
considered. In Ref. \cite{Dukelsky:2001bc} an exactly solvable model of
bosons with a repulsive pairing and single-particle energies
proportional to the angular momentum quantum number was introduced.
This model was further elaborated in Ref. \cite{Pan:2003ki}.
In these models the eigenvalues and the eigenfunctions are written
down in terms of variables that need to be numerically evaluated by
solving non-linear equations. Our aim here is to introduce a model for
which an explicit analytic energy level
expression can be given for all levels and analytic expressions can be
given for at least few lowest eigenstates. A preliminary version of
this work was presented in Ref. \cite{Balantekin:2003zf}.

\section{The Model}

To present our model we first introduce the operator
\begin{equation}
\label{7}
S^+ (\lambda) = \frac{1}{2} \left[ \frac{(d^{\dagger} \cdot
d^{\dagger})}{1
-\lambda} - \frac{s^{\dagger}s^{\dagger}}{\lambda} \right].
\end{equation}
Clearly $S^+{(\lambda})$ is proportional to the operator in Eq.
(\ref{4}) when the parameter
$\lambda = 1/2$. However it represents a general mixing
of $s$ and $d$ bosons when  $\lambda \neq 1/2$.  
Similarly we introduce the operators
\begin{equation}
\label{7a}
S^- (\lambda) = \frac{1}{2} \left[ \frac{( {\tilde d} \cdot
{\tilde d})}{1
-\lambda} - \frac{s s }{\lambda} \right],
\end{equation}
and
\begin{equation}
\label{7b}
S^0 (\lambda) = \frac{1}{4} \left[ \frac{
 \sum_{\mu} (d^\dagger_{\mu}  d_{\mu} + d_{\mu}
d^{\dagger}_{\mu})}{1
-\lambda} - \frac{(s^{\dagger} s + s s^{\dagger})
}{\lambda} \right].
\end{equation}
In this paper we show
that the eigenvalues and the eigenstates of the Hamiltonian
\begin{equation}
\label{8}
{\cal H} = - \frac{1}{4} \left[ \frac{{\hat N}_d}{1
-\lambda} - \frac{{\hat N}_s}{\lambda} \right]^2 +
\frac{1}{2} \left[ \frac{{\hat N}_d}{(1
-\lambda)^2} + \frac{{\hat N}_s}{\lambda^2} \right]
- W(\lambda) \left[ \frac{{\hat N}_d}{1
-\lambda} - \frac{{\hat N}_s}{\lambda} \right]
+ S^+(\lambda) S^-(\lambda) ,
\end{equation}
can be analytically calculated. Here ${\hat N}_d$ and ${\hat N}_s$
are the d- and s-boson number operators, and the function
$W(\lambda)$ is defined below.

The operators given in Eqs. (\ref{1}), (\ref{2}), and (\ref{3})
and their s-boson counterparts satisfy the $SU(1,1)$ commutation 
relations 
\begin{equation}
\label{suii1}
[S^+_i, S^-_j ] = - 2 \delta_{ij} S^0_j,
\end{equation}
and
\begin{equation}
\label{suii2}
[S^0_i, S^{\pm}_j ] = \pm \delta_{ij} S^{\pm}_j,
\end{equation}
where $i,j$, etc. represent different $SU(1,1)$ algebras written using 
only the s,d, etc. bosons. Using Eqs. (\ref{suii1}) and (\ref{suii2}) 
we obtain the commutation relations 
\begin{equation}
\nonumber
[S^+(\lambda),S^-(\mu)] = - 2\frac{S^0(\lambda)-S^0(\mu)}{\lambda-\mu},
\end{equation}
\begin{equation}
\label{GAUDIN_ALGEBRA}
[S^0(\lambda),S^{\pm}(\mu)]=\pm\frac{S^{\pm}(\lambda)-S^{\pm}(\mu)}
                                                  {\lambda-\mu},
\end{equation}
\begin{equation}
\nonumber
[S^0(\lambda),S^0(\mu)]=[S^{\pm}(\lambda),S^{\pm}(\mu)]=0.
\end{equation}
Eqs. (\ref{GAUDIN_ALGEBRA}) were first studied by Gaudin
\cite{Gaudin} in the context of interacting spins on a lattice.
The limit of these commutators when $\lambda = \mu$ can easily
shown to be
\begin{equation}
\nonumber
[S^+(\lambda),S^-(\lambda)] = - 2\frac{\delta S^0(\lambda)}{\delta
\lambda},
\end{equation}
\begin{equation}
\label{GAUDIN_ALGEBRA2}
[S^0(\lambda),S^{\pm}(\lambda)] = \pm
\frac{\delta S^{\pm}(\lambda)}{\delta \lambda},
\end{equation}
Next it is easy to show that the parametric Hamiltonians
\begin{equation}\label{auxH}
H(\lambda)=S^0(\lambda)S^0(\lambda) - \frac{1}{2}S^+(\lambda)
S^-(\lambda) -  \frac{1}{2}S^-(\lambda)S^+(\lambda)
\end{equation}
form a one-parameter family of mutually commuting operators:
\begin{equation}\label{H_COMMUTATORS}
[H(\lambda),H(\mu)]=0.
\end{equation}
Starting from a lowest weight vector and using $J^+(\lambda)$ as
step operators, one can diagonalize them simultaneously. Lowest
weight vector $|0>$ is the boson vacuum:
\begin{equation}\label{LOWEST_WEIGTH_STATE}
S^-(\lambda)|0>=0 .
\end{equation}
Note that our choice of the ground state is different from what is
normally taken in the SO(6) limit, which has the maximum population
of the s-bosons.
The function $W(\lambda)$ we introduced earlier in the Hamiltonian
of Eq. (\ref{8}) is obtained by the action of the operator
$S^0(\lambda)$ on the vacuum state:
\begin{equation}
\label{vacuumw}
S^0(\lambda)|0>=W(\lambda)|0>.
\end{equation}
For the s-d boson model $W(\lambda)$ takes the form
\begin{equation}
\label{wdef}
W (\lambda) = \frac{1}{4} \left[ \frac{5}{1-\lambda} -
\frac{1}{\lambda} \right].
\end{equation}
The boson vacuum itself is an eigenvector of $H(\lambda)$:
\begin{equation}
H(\lambda)|0>=\epsilon_0(\lambda)|0>
\end{equation}
with the eigenvalue
\begin{equation}
\epsilon_0(\lambda)=W(\lambda)^2-W'(\lambda).
\end{equation}
It can be shown that a vector of the form
\begin{equation}
|\xi>\equiv S^+(\xi)|0>
\end{equation}
is also an eigenvector of $H(\lambda)$ if $\xi$ the solution of
\begin{equation}
W(\xi)=0.
\end{equation}
In this case the eigenvalue is given by
\begin{equation}
\epsilon_1(\lambda)= \epsilon_0(\lambda) -
2\frac{W(\lambda)}{\lambda-\xi}.
\end{equation}

Using the algebra given in
Eqs. (\ref{GAUDIN_ALGEBRA})
it is easy to show that the state
\begin{equation}
\label{eigenstate}
|\xi_1,\xi_2,\dots,\xi_n>\equiv S^+(\xi_1) S^+(\xi_2)\dots
S^+(\xi_n)|0>
\end{equation}
is an eigenvector of $H(\lambda)$ if the quantities
$\xi_1,\xi_2,\dots$ satisfy the following system of equations:
\begin{equation}
\label{constraint}
W(\xi_\alpha)=\sum_{\beta\neq\alpha}^n
\frac{1}{\xi_\alpha-\xi_\beta} \quad \mbox{for} \quad
\alpha=1,2,\dots,n.
\end{equation}
The eigenvalue of the Hamiltonian in Eq. (\ref{auxH}) on the
state of Eq. (\ref{eigenstate}) is given by
\begin{equation}
\epsilon_n(\lambda)= \epsilon_0(\lambda)-2\sum_{\alpha=1}^n
\frac{W(\lambda)-W(\xi_\alpha)}{\lambda-\xi_\alpha}.
\end{equation}
Comparing the Hamiltonians of Eq. (\ref{8}) and (\ref{auxH})
and using Eqs. (\ref{GAUDIN_ALGEBRA2}), it follows that
the energy eigenvalues of the Hamiltonian ${\cal H}$ are
given by
\begin{equation}
\label{energy}
E_n(\lambda)= 2\sum_{\alpha=1}^n
\frac{W(\lambda)-W(\xi_\alpha)}{\lambda-\xi_\alpha}.
\end{equation}
All the above considerations are equally applicable to both real 
and complex values of $\lambda$. However only when $\lambda$ is real 
then our Hamiltonian is Hermitian with real eigenvalues. This is 
true even when the solutions of Eq. (\ref{constraint}) are complex.

\section{Calculation of the Eigenvalues and the Eigenfunctions}

In order to evaluate energy eigenvalues and eigenfunctions 
we will use the general expression
\begin{equation}
\label{wdefnew}
W (\lambda) =  \left[ \frac{A}{a_1 - \lambda} +
\frac{B}{a_2 - \lambda} \right]
\end{equation}
so that our calculations are equally applicable to the mixture of
other bosons.
Inserting Eq. (\ref{wdefnew}) into Eq. (\ref{energy}) it is easy
to show that
\begin{equation}
\label{energy2} E_n(\lambda)= 2 \sum_{\alpha=1}^n \left[
\frac{A}{(a_1 - \lambda)(a_1 - \xi_\alpha)} + \frac{B}{(a_2 -
\lambda)(a_2 - \xi_\alpha)} \right].
\end{equation}

It turns out that an expression for energy eigenvalues can be obtained
without explicitly calculating the solutions of Eq. (\ref{constraint}).
From Eq. (\ref{constraint}) it is fairly straightforward to show
that
\begin{equation}
\label{firstsum}
\sum_{\alpha=1}^n W(\xi_\alpha) = 0.
\end{equation}
Similarly considering the expression
\begin{equation}
\label{secondsum}
\sum_{\alpha=1}^n \xi_\alpha W(\xi_\alpha) = \sum_{\alpha=1}^n
\left[ \sum_{\beta < \alpha} \frac{\xi_{\alpha}}{\xi_{\alpha}
-\xi_{\beta}} + \sum_{\beta > \alpha}
\frac{\xi_{\alpha}}{\xi_{\alpha}
-\xi_{\beta}} \right]
\end{equation}
and interchanging the dummy indices $\alpha$ and $\beta$ in the
last sum one finds that
\begin{equation}
\label{secondsumprime}
\sum_{\alpha=1}^n \xi_\alpha W(\xi_\alpha) = \sum_{(\alpha , \beta)
{\rm pairs}}^n \frac{\xi_{\alpha} -\xi_{\beta}}{\xi_{\alpha} -
\xi_{\beta}} = \frac{n(n-1)}{2}.
\end{equation}
Using Eqs. (\ref{wdefnew}) and (\ref{firstsum}) one then gets
\begin{equation}
\label{1app}
A \sum_{\alpha=1}^n \frac{1}{a_1 -\xi_{\alpha}}
= - B \sum_{\alpha=1}^n \frac{1}{a_2 -\xi_{\alpha}}.
\end{equation}
Inserting Eq. (\ref{wdefnew}) into Eq. (\ref{secondsumprime}) and
employing the identity
\begin{equation}
\frac{\xi_{\alpha}}{ a - \xi_{\alpha}} = \frac{a}{a- \xi_{\alpha}} -1
\end{equation}
we get
\begin{equation}
\label{2app}
A \sum_{\alpha =1}^n \left( \frac{a_1}{a_1- \xi_{\alpha}} -1 \right)
+ B \sum_{\alpha =1}^n \left( \frac{a_2}{a_2- \xi_{\alpha}} -1 \right)
=  \frac{n(n-1)}{2} .
\end{equation}
Using Eq. (\ref{1app}) in Eq. (\ref{2app}) one then obtains
\begin{equation}
\label{3app}
B (a_2 -a_1) \sum_{\alpha =1}^n \frac{1}{a_2- \xi_{\alpha}}
=  \frac{n(n-1)}{2} + (A + B) n .
\end{equation}
Using Eq. (\ref{1app}), the energy eigenvalues of Eq. (\ref{energy2})
take the form
\begin{equation}
\label{energy3}
E_n(\lambda)= 2 B \left( \frac{1}{a_2 - \lambda} -  \frac{1}{a_1
- \lambda} \right) \sum_{\alpha=1}^n
\frac{1}{a_2 - \xi_{\alpha}}.
\end{equation}
Finally inserting Eq. (\ref{3app}) into Eq. (\ref{energy3}) we obtain
\begin{equation}
\label{energy4}
E_n(\lambda)= \frac{-2}{(a_1-\lambda)(a_2 - \lambda)}
\left[  \frac{n(n-1)}{2} + (A + B) n \right] .
\end{equation}
In our  model we took $a_1=1$, $a_2=0$, $A=5/4$, and $B=1/4$
resulting in
\begin{equation}
\label{energy5}
E_n(\lambda)= \frac{1}{\lambda(1 - \lambda)}
[ n^2 + 2 n ] .
\end{equation}

Even though we did not need explicit solutions of Eq. (\ref{constraint})
to evaluate the energy eigenstates, we nevertheless need to find these
solutions to be able to write down the energy eigenstates. To achive
this goal we first introduce a change of variables
\begin{equation}
\label{newvariables}
\xi_{\alpha} = a_2 + \zeta_{\alpha} (a_1 - a_2).
\end{equation}
Using these new variables the Bethe ansatz equation Eq.
(\ref{constraint}) takes the form
\begin{equation}
\label{constraint2} \sum_{ \beta\neq\alpha }^n
\frac{1}{\zeta_\alpha-\zeta_\beta} + \frac{B}{\zeta_{\alpha}}
-\frac{A}{1-\zeta_{\alpha}} = 0 \quad \mbox{for} \quad
\alpha=1,2,\dots,n,
\end{equation}
where $B=1/4$ and $A=5/4$ in the s-d boson problem. In Ref.
\cite{stiel}, Stieltjes had shown that the polynomial
\begin{equation}
p_n(\xi) = \prod_{\alpha =1}^{n} (\zeta - \zeta_{\alpha})
\end{equation}
satisfies the hypergeometric differential equation
\begin{equation}
x (1 -x) p_n''(x) + [ 2B -2(A+B)] p_n'(x) + n(n+2A +2B -1)
p_n (x) = 0.
\end{equation}
Hence $\zeta_{\alpha}$ are the roots of the polynomials
\begin{equation}
\sum_k \frac{( -n)_k (n+2A+2B-1)_k}{(2B)_k k!} x^k,
\end{equation}
where
\begin{equation}
(a)_n = a (a+1) \cdots (a+n-1)
\end{equation}
for $n=1,2,\cdots$ and $(a)_0$ = 0. So for example the first
excited eigenstate is
\begin{equation}
\label{firststate}
\frac{A+B}{2} \left[ \frac{(d^{\dagger} \cdot
d^{\dagger})}{A} - \frac{s^{\dagger}s^{\dagger}}{B} \right] |0> ,
\end{equation}
and the second eigenstate is
\begin{equation}
\label{secondstate}
(A+B+1)^2 \left[ \frac{(d^{\dagger} \cdot
d^{\dagger})}{2A+1 - \gamma} - \frac{s^{\dagger}s^{\dagger}}{2B+1 
+ \gamma} \right] \left[ \frac{(d^{\dagger} \cdot
d^{\dagger})}{2A+1 + \gamma} - \frac{s^{\dagger}s^{\dagger}}{2B+1 
- \gamma} \right] |0> ,
\end{equation}
where
\begin{equation}
\label{secondstateaux}
\gamma = \sqrt{\frac{(2A+1)(2B+1)}{2A+2B+1}}.
\end{equation}
Finding the eigenstates of the third, fourth, etc.
excited states simply requires finding the roots of a
cubic, quartic, etc. polynomial. Even though 
for the third and fourth 
order polynomials this can be done analytically, the resulting 
expressions are not particularly illuminating. Of course the roots 
can be calculated numerically for any order.   

\vskip 1cm

This  work   was supported in  part  by   the  U.S.  National
Science Foundation Grants No.\ INT-0352192 and PHY-0244384 at the
University of  Wisconsin, and  in  part by  the  University of
Wisconsin Research Committee   with  funds  granted by    the
Wisconsin Alumni  Research Foundation. The research of YP was 
supported in part by the Turkish Scientific and Technical 
Research Council
(T{\"U}B{\.I}TAK).

\end{document}